\documentclass[journal]{IEEEtran}
\usepackage{cite}
\usepackage{amsmath,amssymb,amsfonts}

\usepackage{mathtools}

\usepackage{tabularx, booktabs}
\usepackage{multirow, makecell}

\usepackage{graphicx}
\graphicspath{{./Figures/}}

\usepackage{subfigure}

\usepackage{array}
\usepackage[ruled,vlined]{algorithm2e}
\usepackage{enumerate}
\usepackage{color}
\usepackage{setspace}

\makeatletter
\let\NAT@parse\undefined
\makeatother

\usepackage[colorlinks=true, linkcolor=blue, urlcolor=blue, citecolor=blue]{hyperref}

\definecolor{Blue1}{rgb}{0.3,0,1}

\newtheorem{remark}{Remark}

\begin{document}

\title{Sample-Efficient Learning for a Surrogate Model of Three-Phase Distribution System}
\author{Hoang~Tien~Nguyen,~\IEEEmembership{Member,~IEEE,} Young-Jin~Kim,~\IEEEmembership{Senior Member,~IEEE,}
	and~Dae-Hyun~Choi,~\IEEEmembership{Member,~IEEE}
	}

\maketitle

\begin{abstract}
A surrogate model that accurately predicts distribution system voltages is crucial for reliable smart grid planning and operation.
This letter proposes a fixed-point data-driven surrogate modeling method that employs a limited dataset to learn the power-voltage relationship of an unbalanced three-phase distribution system.
The proposed surrogate model is designed using a fixed-point load-flow equation, and the stochastic gradient descent method with an automatic differentiation technique is employed to update the parameters of the surrogate model using complex power and voltage samples.
Numerical examples in IEEE 13-bus, 37-bus, and 123-bus systems demonstrate that the proposed surrogate model can outperform surrogate models based on the deep neural network and Gaussian process regarding prediction accuracy and sample efficiency.
\end{abstract}


%
\IEEEpeerreviewmaketitle

\section{INTRODUCTION}
\IEEEPARstart{T}{he} wide deployment of distributed energy sources, such as photovoltaic (PV) systems and electric vehicles, can cause voltage violations in distribution systems. Voltage violations occur due to strongly fluctuating power injections of distributed energy sources, thereby leading to power outages, equipment damage, and customer dissatisfaction.
Therefore, accurate calculation of voltages is essential for various smart grid applications to ensure reliable grid planning and operation.

Distribution system operators have used voltage information to maintain stable system operation by keeping voltage profiles within their acceptable limits via voltage regulators. In addition, from the planning perspective of distribution systems, voltages information has been employed to quantify the influence of newly installed power system equipment on distribution system operations.
Voltage is typically calculated using a model-based load-flow analysis. However, the load-flow analysis requires accurate distribution system models, which are unavailable to most utility companies. Furthermore, distribution system modeling is often costly and inaccurate due to incomplete information about the distribution systems, including topology and transformer/line parameters~\cite{navarro2015reconstruction}.

To address these issues, data-driven surrogate models using machine learning methods have recently been developed to calculate voltage given power injections, without an accurate physics-based system model.
These surrogate models were built using Gaussian process (GP)~\cite{Cao2021Data} and the deep neural network (DNN)~\cite{Cao2022Model}  to approximate the relationship between voltage magnitudes and power injections of the distribution system.
The surrogate models were used as an environment for training process of deep reinforcement learning-based voltage control.
In~\cite{bassi2022electrical}, a DNN-based surrogate model was presented to calculate voltage in low-voltage distribution systems using the historical smart meter data.
In~\cite{balduin2020evaluating}, various machine learning methods were applied and compared to build surrogate models of residential/industrial/commercial distribution systems.
However, these surrogate models suffer from sample inefficiency because they are implemented using purely model-free machine learning methods.

This letter proposes a sample-efficient learning-based surrogate model that learns the underlying power-voltage relationship of an unbalanced three-phase distribution system.
The proposed surrogate model is implemented using a fixed-point load-flow formulation to improve the sample efficiency. During learning, the parameters of the surrogate model are updated using a stochastic gradient descent (SGD) method with an automatic differentiation technique.
Simulation results in IEEE 13-bus, 37-bus, and 123-bus systems demonstrate the superiority of the proposed surrogate model over the DNN and GP methods regarding sample-efficiency and prediction accuracy.
Table~\ref{tab:nomenclature} presents the notations used in this letter.

\begin{table}[t]\footnotesize
	\centering
	\caption{Nomenclature}
	\label{tab:nomenclature}
	\begin{tabular}{ll}
		\toprule
		$N$ & Number of PQ buses \\
		$i \in \{1,2,\dots, N\}$ & Index of a PQ bus \\
		$j$ &Index of a training sample \\
		$k \in \{1,2,\dots, T\}$ &Index of a fixed-point iteration \\
		$\mathbf{v}_0 = (v_0^a, v_0^b, v_0^c)^\top$ & Complex voltages at the slack bus \\
		$\mathbf{v}_i = (v_i^a, v_i^b, v_i^c)^\top $ & Phase-to-ground voltages at bus $i$ \\
		$\mathbf{s}_i^Y = (s_i^a, s_i^b, s_i^c)^\top$ & Grounded wye sources at bus $i$ \\
		$\mathbf{s}_i^\Delta = (s_i^{ab}, s_i^{bc}, s_i^{ca})^\top$ & Delta sources at bus $i$ \\
		$\mathbf{v} = (\mathbf{v}_1^\top, \dots, \mathbf{v}_N^\top)^\top$ &  Complex voltages at PQ buses\\
		$\mathbf{s}^Y = ((\mathbf{s}_1^Y)^\top, \dots, (\mathbf{s}_N^Y)^\top)^\top$ & Complex wye sources at PQ buses \\
		$\mathbf{s}^\Delta = ((\mathbf{s}_1^\Delta)^\top, \dots, (\mathbf{s}_N^\Delta)^\top)^\top$ & Complex delta sources at PQ buses \\
		$\mathbf{Y}$ & Admittance matrix \\
		$\overline{(\cdot)}$ & Conjugate of $(\cdot)$ \\
		\bottomrule
	\end{tabular}
\end{table}

\section{Load-flow formulation of distribution system}\label{sec:problem}

We consider an unbalanced three-phase distribution network with one slack bus and $N$ PQ buses with an admittance matrix $\mathbf{Y} \in \mathbb{C}^{{3(N+1)} \times 3(N+1)}$.
For load-flow analysis, the admittance matrix is partitioned into four block matrices as follows~\cite{Bazrafshan2018Comprehensive}:
\begin{equation}
	\mathbf{Y} \coloneqq
	\begin{bmatrix}
		\mathbf{Y}_{LL} & \mathbf{Y}_{L0} \\
		\mathbf{Y}_{0L} & \mathbf{Y}_{00}
	\end{bmatrix},
\end{equation}
where $\mathbf{Y}_{LL} \in \mathbb{C}^{3N \times 3N}$, $\mathbf{Y}_{L0}\in \mathbb{C}^{3N \times 3}$, $\mathbf{Y}_{0L} \in \mathbb{C}^{3 \times 3N}$, and $\mathbf{Y}_{00} \in \mathbb{C}^{3 \times 3}$ are submatrices of the admittance matrix $\mathbf{Y}$.
\textcolor{black}{The complex voltage $\mathbf{v} \in \mathbb{C}^{3N}$ of the three-phase distribution system can be calculated using the following load-flow equation based on a fixed-point interpretation of the nonlinear AC load-flow equation~\cite{Bernstein2018Load}:}
\begin{align}\label{eq:fixed-point equation}
	\mathbf{v} &= \mathbf{w} + \mathbf{X} \big(\text{diag}(\overline{\mathbf{v}})^{-1} \overline{\mathbf{s}}^Y + \mathbf{H}^\top \text{diag}(\mathbf{H} \overline{\mathbf{v}})^{-1} \overline{\mathbf{s}}^\Delta \big) \nonumber \\
	&=\mathbf{w} + f_\mathbf{X} (\mathbf{v}, \mathbf{s}),
\end{align}
where $\mathbf{X} = \mathbf{Y}_{LL}^{-1}$ and $\mathbf{w} = - \mathbf{Y}_{LL}^{-1} \mathbf{Y}_{L0} \mathbf{v}_0$.
The complex wye and delta sources ($\mathbf{s}^Y, \mathbf{s}^\Delta \in \mathbb{C}^{3N}$) at the buses are $\mathbf{s} = \{ \mathbf{s}^Y, \mathbf{s}^\Delta \}$.
The matrix $\mathbf{H}$ is a $3N \times 3N$ transformation block-diagonal matrix with all diagonal blocks of $[1~ -1~ 0; 0~ 1~ -1; -1~ 0~ 1]$.
Based on the fixed-point equation~\eqref{eq:fixed-point equation}, the complex voltage can be obtained via the following equation:
\begin{equation}\label{eq:fixed-point iteration}
	\mathbf{v}^{k+1} = \mathbf{w} + f_\mathbf{X} (\mathbf{v}^k, \mathbf{s}),
\end{equation}
where the iterations~\eqref{eq:fixed-point iteration} end after $T$ when a change in the voltage between two consecutive iterations is lower than a small threshold.

Note that $\mathbf{w}$ and $\mathbf{X}$ in~\eqref{eq:fixed-point equation} are associated with the distribution system topology and its parameters, which are often not known exactly or are unknown in practice.
This study aims to learn $\mathbf{w}$ and $\mathbf{X}$ using a limited set of complex power and voltage samples $\{\mathbf{s}^{(j)}, \mathbf{v}^{(j)}\}$ collected through sensors.

\begin{figure}[t]
	\centering
	\includegraphics[width=2.5in]{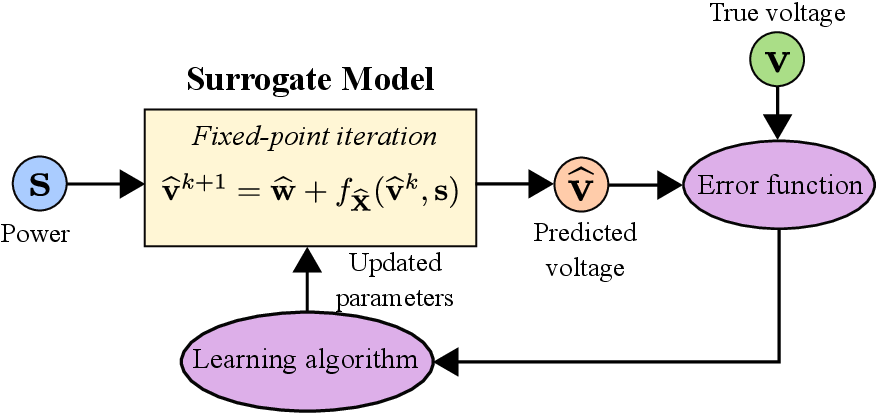}
	\caption{Illustration of a fixed-point surrogate model.}
	\label{fig:surrogate model}	
\end{figure}

\section{Learning-based surrogate model}
\subsection{Proposed Fixed-Point Surrogate Model}

Based on the load-flow equation~\eqref{eq:fixed-point equation}, we propose the following fixed-point surrogate model~\eqref{eq:fixed-point model} to learn the power-voltage relationship of the three-phase distribution system:
\begin{align}\label{eq:fixed-point model}
	\mathbf{\widehat{v}} =\mathbf{\widehat{w}} + f_\mathbf{\widehat{X}} (\mathbf{\widehat{v}}, \mathbf{s}),
\end{align}
where $\mathbf{\widehat{w}} \in \mathbb{C}^{3N}$ and $\mathbf{\widehat{X}} \in \mathbb{C}^{3N \times 3N}$ are model parameters that we need to learn. 
Given complex power $\mathbf{s}$, the predicted complex voltage $\mathbf{\widehat{v}}$ of the surrogate model can be computed using the following iterative equation:
\begin{equation}\label{eq:fp model iteration}
	\mathbf{\widehat{v}}^{k+1} = \mathbf{\widehat{w}} + f_\mathbf{\widehat{X}} (\mathbf{\widehat{v}}^k, \mathbf{s}).
\end{equation}

The primary goal of this study is to determine the optimal estimates for the model parameters $\mathbf{\widehat{w}}$ and $\mathbf{\widehat{X}}$ by minimizing the following root mean square error (RMSE):
\begin{equation}\label{eq:RMSE}
	\mathcal{L} = \sqrt{\frac{1}{3N} \sum_{i=1}^{N} \Vert \mathbf{v}_i - \mathbf{\hat{v}}_i \Vert^2}.
\end{equation}

\subsection{SGD-based Model Training}~\label{subsec:learning method}

This subsection proposes an SGD-based learning method that updates $\mathbf{\widehat{X}}$ and $\mathbf{\widehat{w}}$ by minimizing the error function $\mathcal{L}$~\eqref{eq:RMSE}.
First, the complex matrix $\mathbf{\widehat{X}}$ can be updated using the SGD method for complex variables in~\eqref{eq:gradient} and~\eqref{eq:sgd}:
\begin{align}
	&	\frac{\partial \mathcal{L}}{\partial \mathbf{\overline{\widehat{X}}}} = \frac{\partial \mathcal{L}}{\partial \mathbf{\widehat{v}}} \frac{\partial \mathbf{\widehat{v}}}{\partial \mathbf{\overline{\widehat{X}}}} + 	\frac{\partial \mathcal{L}}{\partial \mathbf{\overline{\widehat{v}}}} \frac{\partial \mathbf{\overline{\widehat{v}}}}{\partial \mathbf{\overline{\widehat{X}}}},    \label{eq:gradient}\\
	& \mathbf{\widehat{X}} \leftarrow \mathbf{\widehat{X}} - \eta \frac{\partial \mathcal{L}}{\partial \mathbf{\overline{\widehat{X}}}}.	\label{eq:sgd}	
\end{align}
Note that, in machine learning packages (e.g., Pytorch), the gradient~\eqref{eq:gradient} can be computed efficiently using \textit{automatic differentiation}--a powerful tool to automate the calculation of derivatives in complex algorithms and mathematical functions.

Next, according to the fixed-point equation~\eqref{eq:fixed-point equation}, the no-load voltage $\mathbf{w}$ is expressed as $\mathbf{w} = \mathbf{v} - f_\mathbf{X} (\mathbf{v}, \mathbf{s})$.
The no-load voltage is updated using the calculated $\mathbf{\widehat{v}}$ and $\mathbf{\widehat{X}}$ as follows:
\begin{equation} \label{eq:no-load voltage update}
	\mathbf{\widehat{w}} \leftarrow \mathbf{v} - f_\mathbf{\widehat{X}} (\mathbf{\widehat{v}}, \mathbf{s}).
\end{equation}

The proposed fixed-point surrogate model and training procedure with the SGD method are illustrated in Fig.~\ref{fig:surrogate model} and Algorithm~\ref{alg:training}, respectively.

\begin{remark}
	The DNN and GP methods for distribution system surrogate modeling typically require numerous  training data to reach an acceptable accuracy because they consider the distribution system a black box (without information).	
	In contrast, the proposed method employs the load-flow principle in~\eqref{eq:fixed-point equation} to build the desired surrogate model. Due to the knowledge of the load-flow principle, only the necessary model parameters are learned, improving the sample efficiency of the learning algorithm.
\end{remark}

\begin{remark}
	While the DNN- and GP-based surrogate models in~\cite{Cao2021Data, Cao2022Model, bassi2022electrical} calculate only the voltage magnitudes given power injection at nodes, the proposed model~\eqref{eq:fixed-point model} provides a complex voltage consisting of the voltage magnitude and angle. The voltage angle is crucial for assessing phase-angle unbalance in three-phase distribution systems.
\end{remark}

\begin{algorithm}[t] \small \label{alg:training}
	\setstretch{1.3} 
	\SetAlgoLined
	\KwIn{Training dataset $\mathcal{D}$, learning rate $\eta$}
	\KwOut{Optimized $\mathbf{\widehat{X}}$ and $\mathbf{\widehat{w}}$}
	Initialize $\mathbf{\widehat{X}}$ and $\mathbf{\widehat{w}}$; \\
	\For{each epoch}{
		\For{each sample}{
			Sample $\{\mathbf{s}^{(j)}, \mathbf{v}^{(j)}\} \sim \mathcal{D}$; \\
			Forward pass: obtain $\mathbf{\widehat{v}}^{(j)}$ by solving \eqref{eq:fp model iteration} with $\mathbf{s}^{(j)}$;\\
			Backward pass: calculate the gradient $\frac{\partial \mathcal{L}}{\partial \mathbf{\overline{\widehat{X}}}}$ using~\eqref{eq:gradient}; \\
			Update $\mathbf{\widehat{X}}$ and no-load voltage $\mathbf{\widehat{w}}$: \\ $\mathbf{\widehat{X}} \leftarrow \mathbf{\widehat{X}} - \eta \frac{\partial \mathcal{L}}{\partial \mathbf{\overline{\widehat{X}}}}$ using~\eqref{eq:sgd};  \\
			$\mathbf{\widehat{w}} \leftarrow \mathbf{v}^{(j)} - f_\mathbf{\widehat{X}} (\mathbf{\widehat{v}}^{(j)}, \mathbf{s}^{(j)})$ using~\eqref{eq:no-load voltage update};
		}
	}
	\caption{Training a fixed-point surrogate model}
\end{algorithm}

\section{Simulation results}
The performance of the proposed fixed-point surrogate model was quantified using IEEE 13-bus, 37-bus, and 123-bus systems with some PV systems, respectively.
\textcolor{black}{The capacity and locations of PV systems are shown in Table~\ref{tab:PV systems}.}

\begin{table}[t!] \small
	\centering
	\caption{\color{black}{Parameters and locations of PV systems}}
	\label{tab:PV systems}
	\textcolor{black}{\begin{tabular}{ccc}
			\toprule
			System & Capacity & Location \\
			\midrule
			13 bus & 10 kW/12 MVA & 634, 675 \\
			\midrule
			37 bus & 900 kW/910 MVA & 705, 710, 730, 736 \\
			\midrule
			\multirow{2}{*}{123 bus} & 150 kW/160 MVA & 13, 18, 44, 105, 152, 250 \\
			& 250 kW/260 MVA & 34, 70, 105, 116 \\
			\bottomrule
	\end{tabular}}
\end{table}

\begin{table}[t!]\small
	\centering
	\caption{Number of training and testing samples}
	\label{tab:sample number}
	\begin{tabular}{cccc}
		\toprule
		& 13 bus & 37 bus & 123 bus \\
		\midrule
		No. training samples & $40$ & $70$ & $200$ \\
		No. testing samples & $1000$ & $1000$ & $1000$ \\
		\bottomrule
	\end{tabular}
\end{table}

\begin{table}[t!]\small
	\centering
	\caption{\textcolor{black}{Training time (min) of three surrogate models}}
	\label{tab:training time}
	\textcolor{black}{\begin{tabular}{cccc}
		\toprule
		& 13 bus & 37 bus & 123 bus \\
		\midrule
		Proposed & 8.8 & 17.3 & 180.7 \\
		DNN & 4.9 & 12.5 & 40.1 \\
		GP & 2.5 & 4.0 & 8.3 \\
		\bottomrule
	\end{tabular}}
\end{table}	

The admittance matrices of the unbalanced distribution systems were obtained from~\cite{Bazrafshan2018Comprehensive}.
The learning rate was initially set to $\eta=0.5$ for training acceleration and multiplied by 0.1 every 2000 epochs. The proposed surrogate model was trained for 7000 epochs.
The no-load voltage $\mathbf{\widehat{w}}$ and all elements of the matrix $\mathbf{\widehat{X}}$ were initialized with the voltages at the slack bus and $0.1 + 0.1j$, respectively.
The uniformly randomized loads and PV system generation outputs generated datasets by computing the load flow~\eqref{eq:fixed-point equation} and obtaining its voltage solution.
\textcolor{black}{The operating status of voltage regulators was assumed to be fixed.}
Table~\ref{tab:sample number} presents the number of training and testing samples.
The performance of the proposed method was compared with those of the GP~\cite{Cao2021Data} and DNN~\cite{Cao2022Model} methods.
\textcolor{black}{The GP-based surrogate model was trained using the loss function of exact marginal log likelihood along with a learning rate of 0.3 during 10000 epochs.
For the training of the DNN-based surrogate model, the loss function of means square error was used with a learning rate of 0.0001. The batch size and number of epochs were 1 and 10000, respectively. The neural networks having two hidden layers with 128, 256, and 500 neurons in each layer were used for IEEE 13-bus, 37-bus, and 123-bus systems, respectively.
Moreover, AdamW optimizer and SGD were employed for training the DNN- and GP-based surrogate models, respectively.
}
For a fair sample-efficiency comparison, identical training and testing datasets were employed for the three methods.
All simulations were performed using PyTorch library in Python.
\textcolor{black}{The simulation study was carried out on a computer with an Intel i7-13700K CPU and an NVIDIA GeForce RTX 4070Ti GPU.}

\begin{table}[t!] \small
	\centering
	\caption{Root mean square error of voltage magnitude using testing dataset}
	\label{tab:rmse}
	\begin{tabular}{cccc}
		\toprule
		& 13 bus & 37 bus & 123 bus \\
		\midrule
		Proposed & $\bf{5.1571 \times 10^{-6}}$ & $\bf 5.3389 \times 10^{-6}$ & $\bf 2.5064 \times 10^{-7} $ \\
		DNN & $1.6494 \times 10^{-3}$ & $7.0648 \times 10^{-4}$ & $1.0229 \times 10^{-3}$\\
		GP & $6.6004 \times 10^{-3}$ & $6.5853 \times 10^{-3}$ & $9.1579 \times 10^{-3}$ \\
		\bottomrule
	\end{tabular}
	\vspace*{-8pt}
\end{table}

\begin{figure}[t!]
	\centering
	\includegraphics[width=2.5in]{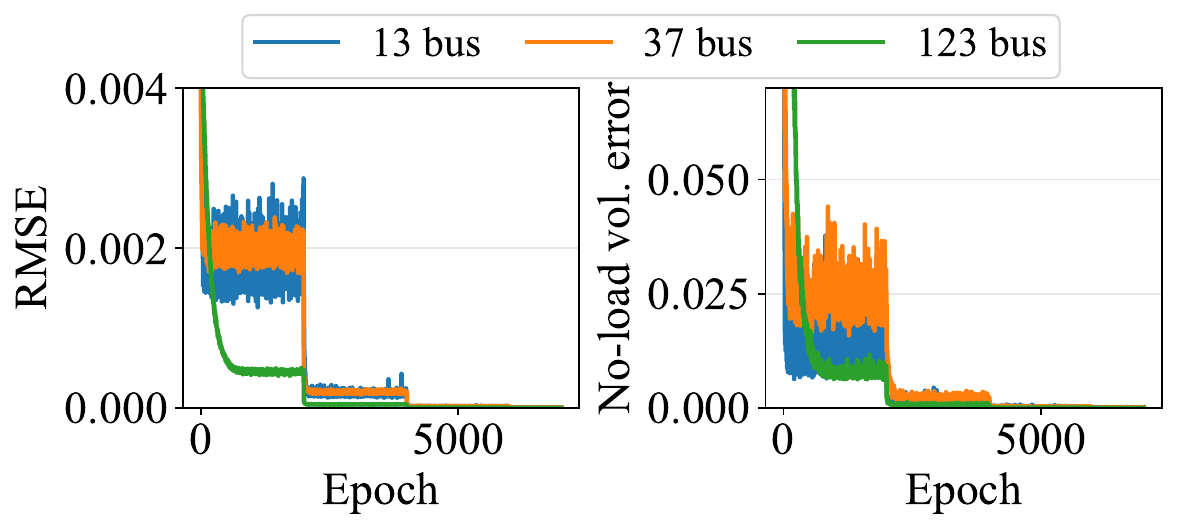}
	\caption{Training curve convergence of the error function $\mathcal{L}$ and no-load voltage error $\Vert \mathbf{\widehat{w}} -  \mathbf{w} \Vert_2 $.}
	\label{fig:training curve}
\end{figure}

\begin{figure}[t!]
	\centering
	\includegraphics[width=2.9in]{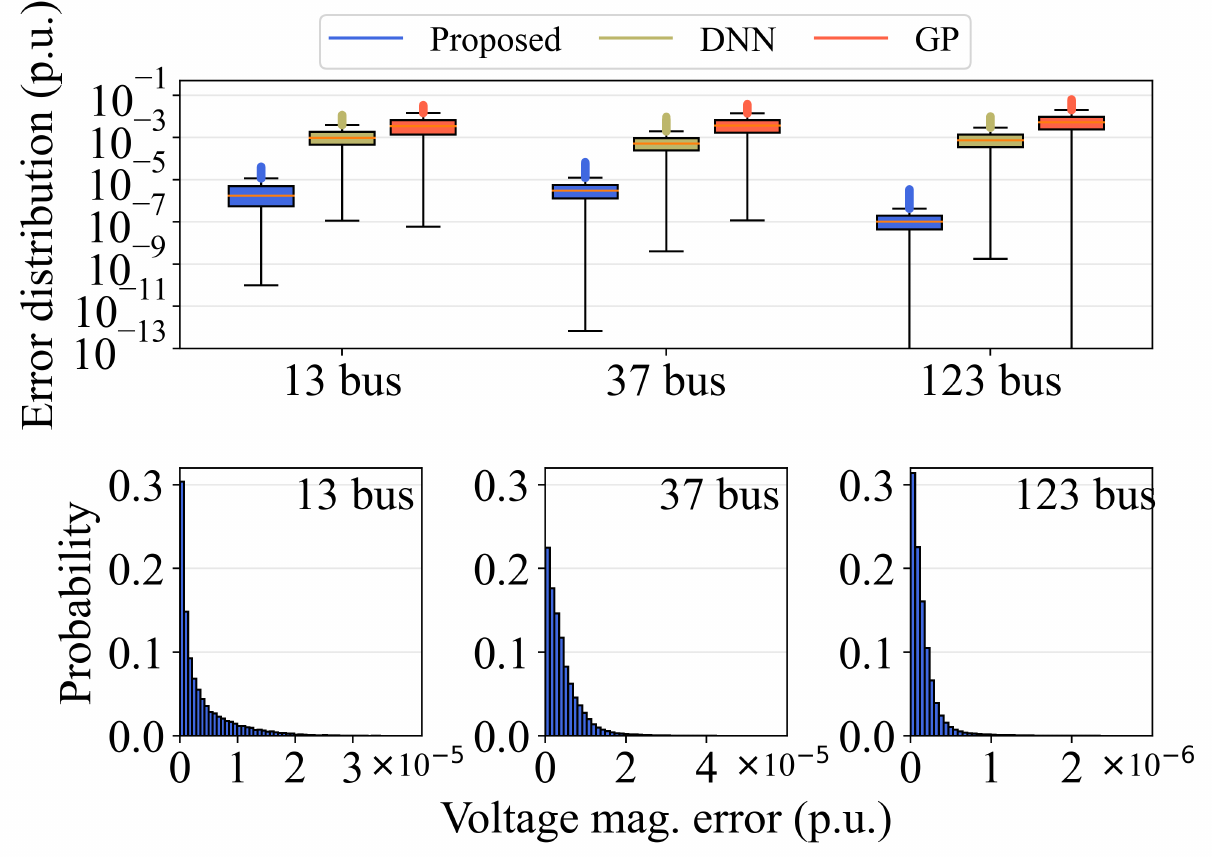}
	\caption{Distributions of voltage magnitude errors of three methods (top) and \textcolor{black}{probabilities of voltage magnitude errors of the proposed method (bottom).}}
	\label{fig:error distribution}
\end{figure}

\textit{Training results:} The training curves of the error function~\eqref{eq:RMSE} and no-load voltage error $\Vert \mathbf{\widehat{w}} -  \mathbf{w} \Vert_2 $ are illustrated in Fig.~\ref{fig:training curve}. This figure reveals that the error function and no-load voltage error converge to near-zero values, demonstrating a successful training convergence of the proposed learning rule for surrogate modeling.
\textcolor{black}{The training times of three methods are shown in Table~\ref{tab:training time}}.

\textit{Testing results:} As listed in Table~\ref{tab:rmse}, the RMSE of voltage magnitude of the proposed method using the testing dataset is much less than those of the DNN and GP methods. Fig.~\ref{fig:error distribution} compares the distribution of voltage magnitude error among the three methods.
The voltage magnitude error of the proposed method is the smallest among the three methods, and its errors are concentrated in a small range near zero.
\textcolor{black}{
Note from Fig.~\ref{fig:error distribution} that the DNN and GP methods generate large voltage magnitude errors during the test phase. This is because the usage of small training sets yields the overfitting phenomenon in the training phase. That is, the DNN- and GP-based surrogate models perform perfectly on the training dataset, while poorly fitting on the test dataset. In practice, the DNN and GP methods need more training datasets to achieve the surrogate model accuracy equivalent to the proposed method. By contrast, the proposed method combined with the fixed-point load-flow equation exploits the training samples more efficiently than the DNN and GP methods.
}

\textcolor{black}{In addition, the RMSE values of the voltage angles were calculated as $3.3631 \times 10^{-6}$, $ 5.2793 \times 10^{-6}$, and $2.5233 \times 10^{-7}$ rad for IEEE 13-bus, 37-bus, and 123-bus systems, respectively.}
\textcolor{black}{The average inference time of three surrogates models is shown in Table~\ref{tab:inference time}. The inference time of DNN is smallest among three models, and the inference time of the proposed model is smaller than that of GP.}
The results confirm high prediction accuracy, sample-efficiency, and fast \textcolor{black}{inference} of the proposed model.

\begin{table}[!] \small
	\centering
	\caption{\textcolor{black}{Inference time (ms) of three surrogate models}}
	\label{tab:inference time}
	\textcolor{black}{\begin{tabular}{cccc}
		\toprule
		& 13 bus & 37 bus & 123 bus \\
		\midrule
		Proposed & 2.4 & 3.3 & 9.1 \\
		DNN & 0.2 & 0.3 & 0.3 \\
		GP & 4.1 & 18.8 & 83.6 \\
		\bottomrule
	\end{tabular}}
\end{table}

\section{Conclusion}

This letter proposes a fixed-point surrogate model in which the power-voltage relationship of the unbalanced three-phase distribution system is learned by applying a limited dataset of complex power and voltage.
The critical parts of the proposed method are to i) design the surrogate model based on the fixed-point load-flow equation and ii) learn the parameters of the surrogate model using the SGD method.
Due to the leverage of the load-flow equation, the proposed surrogate model can achieve higher prediction accuracy and sample efficiency than the DNN- and GP-based surrogate models, as verified in the simulation study.
This study demonstrates that combining domain knowledge (i.e., the load-flow equation) and machine learning methods could yield highly sample-efficient learning algorithms for power system applications.

In a future study, the proposed method will be extended to identify the distribution system topology and line parameters by learning a full admittance matrix.


%
%
\bibliographystyle{IEEEtran}
\bibliography{references}

\end{document}